\documentclass[]{aastex631}

\shorttitle{Developing a spectropolarimeter}
\shortauthors{Soam et al.}
\graphicspath{{./}{figures/}}

\begin{document}

\title{Spectropolarimeter on a 2--4 m class telescope and proposed science cases}

\correspondingauthor{Archana Soam, Siddharth Maharana}
\email{archana.soam@iiap.res.in, sidh345@gmail.com}

\author[0000-0002-6386-2906]{Archana Soam}
\affiliation{Indian Institute of Astrophysics, II Block, Koramangala, Bengaluru 560034, India}
\author[0000-0002-7072-3904]{Siddharth Maharana}
\affiliation{Institute of Astrophysics, Foundation for Research and Technology-Hellas, GR-71110 Heraklion, Greece; Department of Physics, University of Crete, GR-70013 Heraklion, Greece}
\affiliation{South African Astronomical Observatory, PO Box 9, Observatory, 7935, Cape Town, South Africa}
\author[0000-0002-0786-7307]{B-G Andersson}
\affiliation{SOFIA Science Center/USRA, NASA Ames Research Center, M.S. N232-12, Moffett Field, CA 94035, USA}
\affiliation{Visiting Scholar, Institute for Scientific Research, Boston College, Kenny Cottle Hall 217, 885 Centre St., Newton, MA 02459}
\author{Ramaprakash, A. N.}
\affiliation{Inter-University Centre for Astronomy and Astrophysics, Post Bag 4, Ganeshkhind, Pune 411 007, India}



\begin{abstract}
We propose a spectropolarimeter a covering wavelength range of 3200--7000 {\AA} [3200{\AA} chosen as lower limit to go to the atmospheric cut-off. It's ``needed" for some Serkowski curves and would make the instrument even more unique] for a 2-4~m class telescope. In this article, we discuss the science cases which will be covered with this proposed instrument. The technical requirements and analysis plan for each science case is also discussed. This spectropolarimeter targeting exciting galactic and extra-galactic research, will be unique instrument on a 2-4~m facilities.
\end{abstract}



\section{Introduction} \label{sec:intro}
Interstellar magnetic fields can be mapped with the polarization of starlight caused by aligned dust grains in different lines-of-sights and in different environments.
This was first observed in optical wavelengths$^{11,12}$ and was explained to be arising from dichroic extinction by elongated dust grains aligned with the ambient magnetic field. Interstellar polarization also provides an opportunity to study the dust grain characteristics e.g. size distribution and composition. 

An important science to be addressed with a spectropolarimeter is the investigation of interstellar dust. The differential extinction giving rise to the polarization curve (parameterized by the empirical, convex, ``Serkowski function"$^{36}$
\begin{equation}
p(\lambda)=p_{max} \cdot exp[-K \cdot ln^2(\lambda/\lambda_{max}]
\end{equation}
\noindent where $p_{max}$ is the maximum polarization, located at $\lambda_{max}$, and ``K" controls the width of the curve.  As shown by (ref. 17), the Serkowski form can be understood in terms of Mie scattering off an underlying MRN$^{24}$ grain size distribution - if grains smaller than $\sim$0.045$\mu$m are not aligned.
The Serkowski fit to the polarization spectrum of star-light shinning through this interstellar dust can therefore constrain the size of \textbf{aligned} dust grains$^{17}$. Adding extinction information to the amount of polarization provides important complementary information that can break the degeneracy between the underlying grain size distribution, and the size dependent grain alignment mechanisms, and their efficiencies.  Combining spectroscopy and polarimetery is a time efficient and better approach to perform such investigations in comparison to the often used technique of imaging polarimetry with use of filters. 

A spectropolarimeter at a 2--4-m class telescopes is needed to address some very important science cases, especially those related to interstellar dust. Polarimeters, especially with spectroscopic capabilities are not often available on most observatories- imaging cameras and spectrographs are the most commonly available instruments for telescopes. The reason for this trend is two fold: (a) the complexities and challenges in the development and operation of accurate polarimeters and (b) often, the scientific rational for these kind of instruments are not fully appreciated in the astronomy community. 
\par In this white paper, we describe the scientific motivation to develop a spectropolarimeter for 2--4-m class telescope, along with instrument concepts that are feasible and can meet the technical goals required for the science presented in this paper.

\section{Science cases}

\subsection{Interstellar Medium Science}

This proposal for developing a spectropolarimter is primarily motivated by the  Galactic ISM science cases listed below. 
\begin{enumerate}
    \item \textit{Spectropolarimetry of stars in or behind nearby star-forming regions}:
This type of study is needed to obtain the polarization curve for a number of environments and source types.  This provides a direct means to probe the size distribution of ISM dust$^{2}$. It is a powerful technique that for general ISM studies need 2-4 m telescopes, which can use background stars of any spectral class and provides inherently complementary probes to infra-red data. For spatially restricted regions, where the background star selection is severely restricted, requiring a fainter magnitude limit, 4$+$ m telescopes become necessary. As discussed above, the ISM polarization spectrum shows a convex shape with a peak in the range $\lambda_{max}$ = 0.4$-$0.7$\mu$m$^{35, 41}$. Modern “Radiative Alignment Torque (RAT) alignment” theory$^{17}$, explains this relation by predicting that grains are aligned if exposed to an anisotropic radiation field with wavelengths less than the grain diameter. A universal linear relationship between $\lambda_{max}$ and extinction ($A_{V}$) has been found by (ref. 3) .

\begin{equation}
    \lambda_{max} = (0.027\pm0.002)A_{V} + (0.166\pm0.003)R_{V}
\label{AP_eq}
\end{equation}

where $R_{V}$ is the average total-to-selective extinction. This result can be understood as due to the progressive reddening of the radiation field into the clouds, but with a fixed, underlying, total grain size distribution throughout each cloud. The dependence on $R_{V}$ indicates that the average grain size varies from cloud to cloud.  As discussed below in item \ref{sec:grain_growth}, grain growth in the denser parts of the cloud can be traced$^{41}$ from the deviations of the linear relationship in equation \ref{AP_eq}.

    \item \textit{Investigating grain alignment efficiencies in the regions filled with FUV radiation (high intensity radiation fields)}
This project is an extension of our study of the grain alignment variations around the wall of the Local Bubble$^{25}$ and our recently published paper on IC59 and IC63 clouds illuminated by B0 IVe star $\gamma$ Cas$^{38}$. We used our polarization and extinction measurements to estimate polarization efficiencies in these clouds. We investigated how these polarization efficiencies vary with the flux from gamma Cas. Is this UV illumination related to the polarization and hence grain alignment efficiencies in the clouds?  This experiment was performed on IC 59 and IC 63 and we compared our results with other regions. We saw a change in the slope of the in polarization efficiencies as we went toward the higher illumination sides. Since IC 59/63 clouds are relatively closer to their illuminating source as opposed to regions like local bubble, we saw high poloarization efficiencies in these clouds. Figure 1 below shows these results. However, as we move towards the regions with more luminous source(s), such as LDN 204$^{6}$ and especially the Orion Veil, we see that the polarization efficiency seem to saturate (Soam \& Andersson., in prep.). We speculate that this is either because in the clouds closer to the radiation source all the alignable grains have aligned by the radiation, or that as the radiation field becomes very strong the large grains are destroyed through the so called RATD destruction$^{15}$. Because both RAT alignment and RAD Destruction are more efficient for larger grains, but change the polarization efficiency in opposite directions, broad-band spectro-polarimetry would allow the two possibilities to be separated. Our proposed spectropolarimeter is the best suit for that purpose.


\item \textit{Investigation of dust in circumstellar envelopes (CSEs) of AGB stars.}

The circumstellar envelopes of asymptotic giant branch (AGB) stars provide unique mineralogical laboratories for studying the effects on the grain alignment. This is because of the dredge-up of newly synthesized carbon during the thermal pulses of the AGB phase and the stability of the CO molecule.  When the star first ascends the AGB the circumstellar envelope (CSE) is oxygen rich with all the carbon tied up in CO molecules.  The dust is therefore made up of silicate and metal oxide grains.  As the dredge-up of newly processed material is transferred into the CSE, the C/O ratio eventually exceeds unity and all the oxygen is tied up in CO, with the dust now made up from carbon  materials. Thus, the two types of AGB envelopes provide the opportunity to study the silicate and carbon dust separate from each other. As with interstellar medium studies, the Serkowski-curve parameters can yield important information about the dust in the CSE.  Grain formation and its influences on the formation and characteristics of the AGB winds can be studied by probing the light from field stars background to the AGB CSE.  Models of the initiation and mass loss characteristics of these winds are still an unresolved issue. The brightest AGB CSEs can be probed by FIR and sub-mm wave data at high angular resolution. However, for such emission polarimetry the complex interaction of dust temperature, emissivity and column density/opacity can be difficult to disentangle.  With telescopes and polarimeters capable of reaching V~16 in (at least) BVRI polarization curves at several different points in the CSEs can be acquired, probing the alignment efficiency as well as the grain size distribution (at different radii).  In initial studies of the IRC+10216 CSE we see a large $\lambda_{max}$ and K (although the Serkowski form may not be ideal) indicating large grains as compared to the ISM.  This is consistent with the results of ab initio AGB wind models$^{33, 4}$, but better S/N is needed.  We note that, in the intense radiation field from the central star, the RATD grain destruction mechanism should be important in these CSEs,  Measuring spectral types (and thereby the extinction), and polarization of a number of stars shining through the envelopes of the AGB stars can provide observational constrains on the dust grain size in the AGB stars. The fact that our observations indicate large grains without a small grain population indicates that such destruction is \textbf{not} important.  This may indicate grains with high tensile strength and hence compact solid structures.  Again more and better data are needed to fully explore these possibilities. Based on FIR/sub-mm wave maps of the nearest AGB stars and Gaia distance and photometry many of the nearby AGB stars have several (in some cases $>$10) background field stars within the projected 70$\mu$m extent of their CSEs. An extension of such studies would be to probe the ionized regions of planetary nebulae, where the evolved star and dust interacts with the hard UV of the central star.

\item \textit{Grain Growth and Destruction}\label{sec:grain_growth}
The polarization curve, under RAT alignment, is determined by the size distribution of aligned grains and the SED of the radiation field. 
The smallest aligned grain is given by the largest of the smallest grain that satisfies the RAT criterion and smallest grain not disaligned by collisions$^{41}$.
\begin{itemize}
    \item $\lambda_{min} <$d, where $\lambda_{min}$ is the shortest wavelength available and d is the grain diameter: Lyman limit (912{\AA}) for the diffuse, neutral, ISM; shorter for fully ionized gas close to hot stars$^{7}$, or longer for material inside of ISM clouds where reddening is active.
    \item Thermal collisions disalign the smallest grains first.  The collision rate goes at $n*\sqrt(T)$ and the grains disalignment time goes as $\sim$1/d.  Hence for denser, warmer gas the limit of aligned grains grows$^{41}$.
The over-all grain size distribution, and the upper cut-off of it is determined by grain growth and destruction.
    \item In dense gas (clumps in clouds) the density can become large enough that significant grain growth can occur$^{41}$, extending the size distribution and affecting the red part of the polarization curve. 
    \item If significant grain destruction (in ISM shocks, or through the RATD mechanism) take place, the largest grains may be destroyed shifting the polarization curve to the blue$^{9}$.
\end{itemize}
These various effects can be probed by high-quality determination of the polarization curve in different environments.  (ref. 41) studied the polarization curve variations in the Taurus molecular cloud and found evidence for the effects of reddening, gas-dust collisions disalignment and grain growth in clumps dark and cold enough for ISM ices to have formed. This is manifested as a bifurcation in the $\lambda_{max}$ vs. A$_V$ relation (equ. \ref{AP_eq}), such that, beyond A$_V \approx 4$ mag., a second, much steeper relationship was added.  This opacity is approximately that at which CO$_2$ ice forms in the cloud and it may be that grain growth is initiated via the mediation of the collision damping and sticking characteristics of the ice.

For seven of the 28 stars so far probed in space UV polarization, (ref. 7) found that the UV polarization was elevated (and the location of $\lambda_{max}$ shifted to the blue). All of these stars probe the inside the Per OB3 super bubble, and the change in the polarization curve may therefore be due to the dist characteristics, or UV field, inside the super bubble. The high FUV polarization would imply a EUV ($\lambda <$912 \AA) aligning radiation field.  Since EUV radiation cannot propagate through the neutral ISM, the FUV polarization (and supporting optical spectro-polarimetry) provides a unique probe of the EUV SEDs of young stars (and e.g. Wolf-Rayet stars).

(ref. 9) found that stars in the Local Bubble seemed to show a systematically smaller $\lambda_{max}$ that those further away. For the latter two of these studies the relative influence of the aligning SED and grain size changes are not clear.  

Follow up studies of (ref. 41) should probe other clouds, both to verify and extend the study and to probe it in different environments (star formation activity etc.  Preferably, these should also include NIR (H-band) polarimetry for line of sight deeper than $A_{V}\sim$5 mag.  One such study has been initiated on the Southern Coalsack using SALT/SAAO). 
For the type of studies performed by (ref. 7) and (ref. 9) further high-quality polarimetry, in regions of likely shocks and/or hot start nearby, is needed as well as high quality multi-band photometry to allow reliable extinction curve to separate out spectral energy distribution (SED) and grain destruction effects. Such studies can be used to constrain both the local ISM radiation field, grain growth and destruction as well as cloud parameters such as gas density and ice formation onsets. With the availability of Gaia DR3 data, accurate distances, and usually extinction data are available for comparison and calibration.

\item \textit{Estimation of extinctions and polarization to understand the polarization efficiency in starless cores.}
The degree of polarization is found declining beyond extinctions (Av) of $\approx$20. This phenomena is  named as 'polarization hole' in molecular cloud cores specially the ones without internal sources a.k.a. starless/prestellar cores such as L183 (Soam \& Andersson et al., in prep.) $^{1, 16}$ and cloud 109 of the Pipe nebula$^{1}$. Optical spectroscopy of background stars in many such cores are needed to study the variation of polarization with extinction/column density in the outer parts where the column density is too low to allow accurate far infrared observations. That will increase the sample and provide statistical significant results 'polarization hole' and help understand the critical densities beyond which this happens. Polarization efficiencies i.e. a ratio of degree of polarization and extinction (Av) will also be investigated in a number of starless cores to understand the dust grain alignment. 

\item \textit{Dust properties/distribution in high-latitude clouds and high velocity clouds}
Dust properties and total-to-selective ratio (R$_V$) of high-latitude clouds are found to be different from the regions in the Galactic plane. (ref. 28) found that the polarization efficiency in some such clouds can exceed the "canonical value" of p/A$_V\leq 3\%/mag$, by a factor of 2-3.  Such high polarization efficiencies are difficult to reconcile with standard assumptions about interstellar grains$^{13}$.  Dust properties such as size can constrained by performing spectropolarimetry and fitting Serkowski curve to the background stars shining through the diffuse structures seen at high-galactic latitudes. The maximum polarization can be used as an accurate measure of the alignment efficiency.

\item \textit{Spectropolarimetry of Herbig-Haro objects}
Herbig-Haro (HH) objects are regions of shocked gas caused by the interaction of jets from young atsra and the ISM. These objects can be studied using optical spectropolarimetry. Such a study was performed by (ref. 8) on three HH objects. That study was motivated from the findings of (ref. 34) where they presented the unpolarized emission line from the knot of HH24 arose $in-situ$, whereas there was a highly polarized continuum emission in the same knot, which might be the result of reflected light from a hidden star later found to be T Tauri type star. Therefore, it was suggested that optical spectropolarimetry can be a unique tool to understand the origin of HH objects and identifying the type of stars which are capable of ionising/shocking the nearby condensations. We can survey the HH objects for such investigations using proposed spectro-polarimeter. 

\subsection{Galactic Stellar Science}

While the spectropolarimetry proposal is primarily built around ISM science, several areas of stellar and extra-galactic research could also benefit from these capabilities.  Here we give a small number of examples

\item \textit{Spectropolarimetry of white dwarfs:}
(ref. 43) presented a study on optical polarimetric survey of a number of white dwarfs (WDs) to (i) to perform a statistical analysis of the linear polarization properties of WDs; and (ii) to provide observers with new faint linear polarimetric standard sources. That study was performed on sources whose spectral types are known. Out of the currently know $\sim$23,000 WDs, most of the sources have spectral type information, but there still a number of WDs with no spectral information. Using the proposed spectropolarimeter, we can conduct a similar survey as that of Zejmo et al. on a larger number of WDs with unknown spectral types. 

\item \textit{Spectropolarimetry of Magnetic Chemically Peculiar (CP) Stars:}
(ref. 35) recently published a study on spectropolarimetric survey of 56 chemically peculiar (CP) stars in the association of Orion OB1. Such studies are necessary to to understand the effect of the magnetic field on the evolution of CP stars. The field strength is found to be dropping sharply with the age. The regions line Orion are chosen for such studies as they harbor a plenty of young stars. The specific orientation and spatial motions of the stars in such regions can be used to trace the birthplaces of these stars. Spectropolarimeteric studies of such CP stars may help in understating the evolution of the regions such as Orion and others. 

\end{enumerate}

\begin{table}[]
    \centering
    \caption{Scientific and technical Requirements}
    \begin{tabular}{|c|c|c|}
    \hline
    Science Cases & class,type, number of object & Technical Requirements Checklist \\
    \hline
    Nearby star-forming regions and  & 15-16 V mags & Broadband filters  needed:yes\\
    envelopes of AGB stars && \\
    && Spectropolarimetry needed:Yes,\\
   && Wavelength Coverage:U BVRI\\
  && Spectral Resolution: 100 - 300\\
  && Pol Accuracy: ($<\sigma_p>$): 0.1\\
  && FoV: 10 arcmin \\
  \hline
Investigating grain alignment efficiencies & 15-16 V mags &  Broadband filters  needed:yes\\
    && Spectropolarimetry needed:Yes,\\
   && Wavelength Coverage:U BVRI\\
  && Spectral Resolution: 100 - 300\\
  && Pol Accuracy: ($<\sigma_p>$): 0.1\\
  && FoV: 10 arcmin \\
  \hline
Grain Growth and destruction: Supernovae,& 15-16 V mags&   \\
regions filled with FUV radiation & &  Broadband filters  needed:yes\\
    && Spectropolarimetry needed:Yes,\\
   && Wavelength Coverage:U BVRI\\
  && Spectral Resolution: 100 - 300\\
  && Pol Accuracy: ($<\sigma_p>$): 0.1\\
  && FoV: 10 arcmin \\
\hline
    \end{tabular}
    \label{tab:1}
\end{table}

\subsection{Extra-galactic studies}
This effort of proposing a low-resolution spectropolarimeter is based on the Galactic ISM science cases described above. However, just as with stellar astrophysics, there are number of extra-galatic science cases which can be addressed using this instrument. Some of them are listed below. 
\begin{enumerate}
\item \textit{Spectropolarimetry of type Ia supernovae}
In a study by (ref. 29), the attempt was made to investigate the properties of extra-galactic dust by spectropolarimetry of four Type Ia Supernovae in other Galaxies. The anomalous extinction curves of these Supernovae were analyzed to understand the environment in which they occurred. The authors found an exceptionally low Rv value and an anomalous behavior of the continuum polarization curve. Type Ia supernovae are very luminous and have a negligible intrinsic continuum polarization. One possible explanation of the low R$_V$ values and the progression of polarization curves, with successively smaller $\lambda_{max}$ values is that the RATD dust destruction mechanism is changing the host-galaxy dust distribution responsible for the polarization.  Hence, tracking the polarization curve evolution of bright supernovae over the first few weeks after explosion could provide unique information on the dust characteristics and amount in the host galaxies.
\item \textit{Spectropolarimetry of Active Galactic Nuclei (AGNs) and Blazars}
Spectropolarimetry of AGNs is a unique tool to study the dust around them. (ref. 26) presented the importance of spectropolarimetric surveys on AGNs and their implications for the existence of two populations of type 2 Seyfert galaxies — hidden Seyfert 1s and “true” Seyfert 2s. Such surveys can be performed with the proposed instrument. These investigations will also be useful in detecting double-peaked H$\alpha$ emission in the polarized-flux spectrum of AGNs as was detected in NGC 2110. 
Blazars, which a class of AGNs with a relativistic jets with continuum linear polarization that is typically of a few percent, can also be investigated using proposed spectropolarimeter. Optical linear polarization percent and position angles can help in determining how the head-on orientation or the presence of a jet influences spectropolarimetric variations of blazars in the broad lines and continuum.

\end{enumerate}

\section{Telescope and Instruments}

In general, polarimetry is considered challenging and a niche area within astronomy due to the following reasons:
\begin{itemize}
    \item Most sources, especially stars, emit in optical wavelengths with low polarization signals of 1~\% or less$^{12}$ in $p$. Thus, polarimeters need to be designed to achieve accuracies of 0.1~\% or better.
    \item Telescope and optical elements inside an instrument introduce polarization, which leads to the polarization of the beam getting modulated and modified. To estimate the true polarization of a source with accuracies of 0.1~\% or better, the polarization behaviour of the instrument has to be carefully modelled and corrected during operational lifetime. Most often, for narrow FoV polarimeters, this is done by observing ``polarimetric standard" stars on sky$^{31, 32, 5}$. For wide FoV instruments, dedicated and more involved calibration methods are required$^{10, 21, 22}$.
    \item High accuracy polarimetry requires finding small differences between large numbers (intensities) and is a very photon-hungry method in comparison to photometry and spectroscopy. The relationship between polarimetric accuracy (based only on photon noise) and photometric SNR is given by Equation~\ref{snr_eqn}. As can be seen, achieving a $\sigma_p =  0.1~\%$ requires a photometric SNR of 1410, requiring substantial amount of telescope time in comparison to photometry. 
\begin{equation}\label{snr_eqn}
    \sigma_p = \frac{\sqrt{2}}{SNR_{phot}}
\end{equation}
\end{itemize}

\begin{figure}
    \centering
    \includegraphics[scale = 0.35]{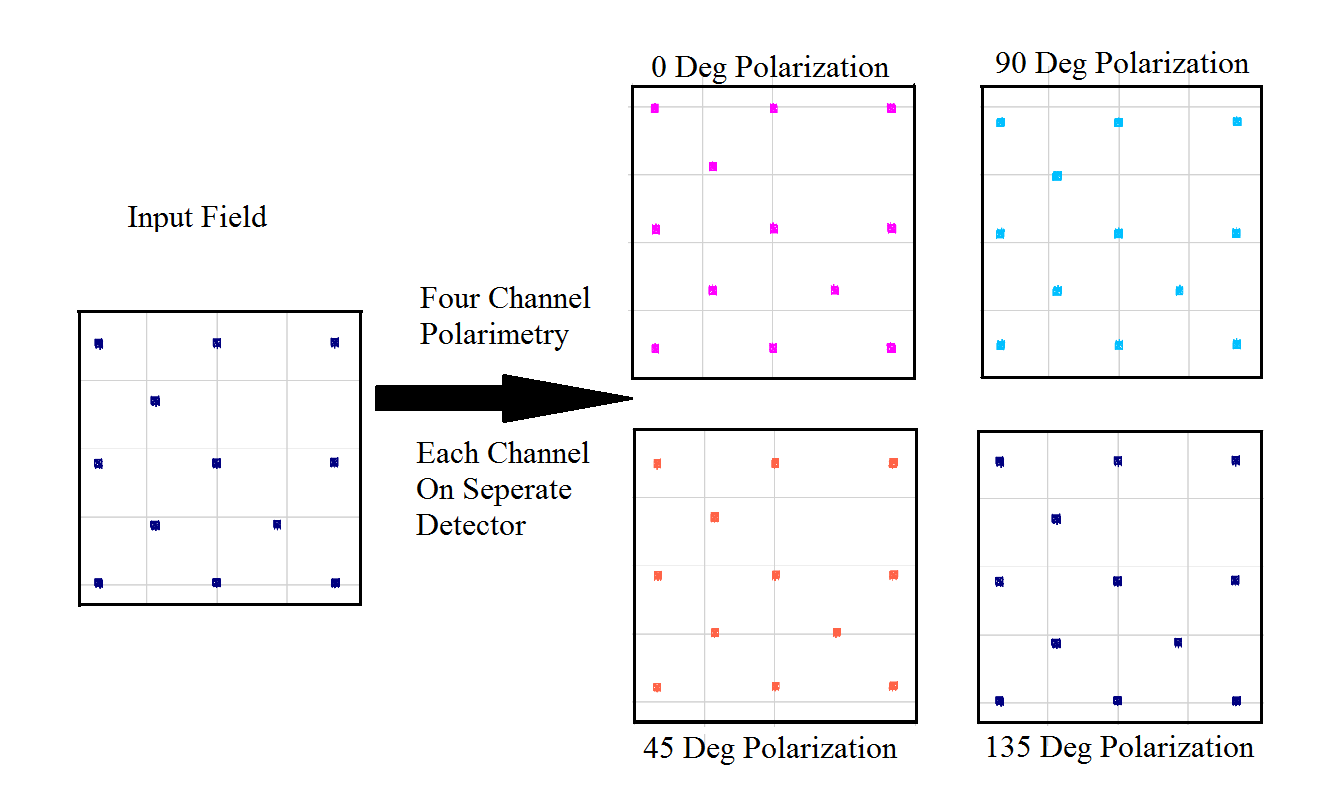}
    \caption{The image is adopted from (ref. 20). It shows the concept of four channel imaging with separate camera for each channel employed in the WALOP instruments. The input field is split into four channels along the $0^{\circ}$, $45^{\circ}$, $90^{\circ}$ and $135^{\circ}$ polarization angles and imaged on four
    separate detectors without any overlap.}
    \label{four_camera_concept}
\end{figure}

Most polarimeters built till date have narrow FoV of a few arcminutes or less, mostly allowing for point source polarimetry. New generation polarimeters are being built which will have capability of wide-field imaging polarimetry such as Wide-Area Linear Optical Polarimeters (WALOPs)$^{20}$ for the PASIPHAE survey$^{19, 39}$. The design and calibration of our proposed spectropolarimeter will be based on the methods developed for the WALOP instruments. The top level concept of WALOP instrument is shown in Figure~\ref{four_camera_concept}, taken from (ref. 20). WALOPs employ the concept of four-channel and four-camera  one-shot polarimetry, whereby each of the polarized beams (channels) coming out of the polarization analyzer system is steered and imaged into its own camera detector. This technique has the following three major advantages over conventional polarimetry:
\begin{enumerate}
    \item Polarimeters in astronomy usually employ two-channel polarimetry, which requires multiple exposures to be taken at various orientations of the analyzer system to estimate the two linear Stokes parameters $q$ and $u$. During any such observation sequence, changing sky conditions as well moving/rotating optical components can introduce measurement errors and lower the instrument's accuracy. To overcome this, one-shot polarimetry has been devised that allows both the linear Stokes parameters to be obtained in one-shot, as demonstrated by the success of RoboPol polarimeter$^{31}$.
    \item  Unlike stars and other point sources, the sky background in an instrument's field of view is an extended object. While the light from the source is split (and divided roughly) by a factor of two or four (= number of channels) by the analyzer, only when the channels are imaged on different detectors/detector areas, the sky background is reduced by the same factor by avoiding overlap of ordinary and extraordinary images from adjacent sky regions. 
    \item There is no overlap of images or spectra from different channels, which will enable accurate photometry and polarimetry in sky regions with higher stellar density. 

    
\end{enumerate}

\begin{table}[t]
    \centering
    \begin{tabular}{cc}
    \hline
    \textbf{Parameter}     & \textbf{Requirement}  \\
    \hline
     Polarimetry   & Linear Polarimetry, Possibly Circular Polarimetry \\
    FoV & $5\times5$~arcminutes or more\\
    Polarimetric Accuracy& 0.1~\% \\
    Wavelength Range &  320-800~nm \\
    Spectropolarimetry & Yes\\
    Imaging Polarimetry & Yes \\
    Spatial Resolution & Seeing Limited (1" full-width half maximum) \\ 
    Spectral Resolutions & 100, 200, 300 \\
    Broadband Filters & U, B, V, R, I \\
     & Two-channel and two-camera four-shot polarimetry\\
     Type of polarimetry & OR\\
     & four-channel and four-camera one-shot polarimetry\\
     \hline
    \end{tabular}
    \caption{Technical requirements for the instrument based on the science case requirements. }
    \label{inst_spec_table}
\end{table}

\subsection{Case for a 4~m telescope}

\begin{figure}[t]
\caption{Polarimetric accuracy vs. exposure plots for different visual brightness of stars observed with 3.6m telescope.}
\includegraphics[trim = {7cm 0cm 7cm 0cm}, width=14cm]{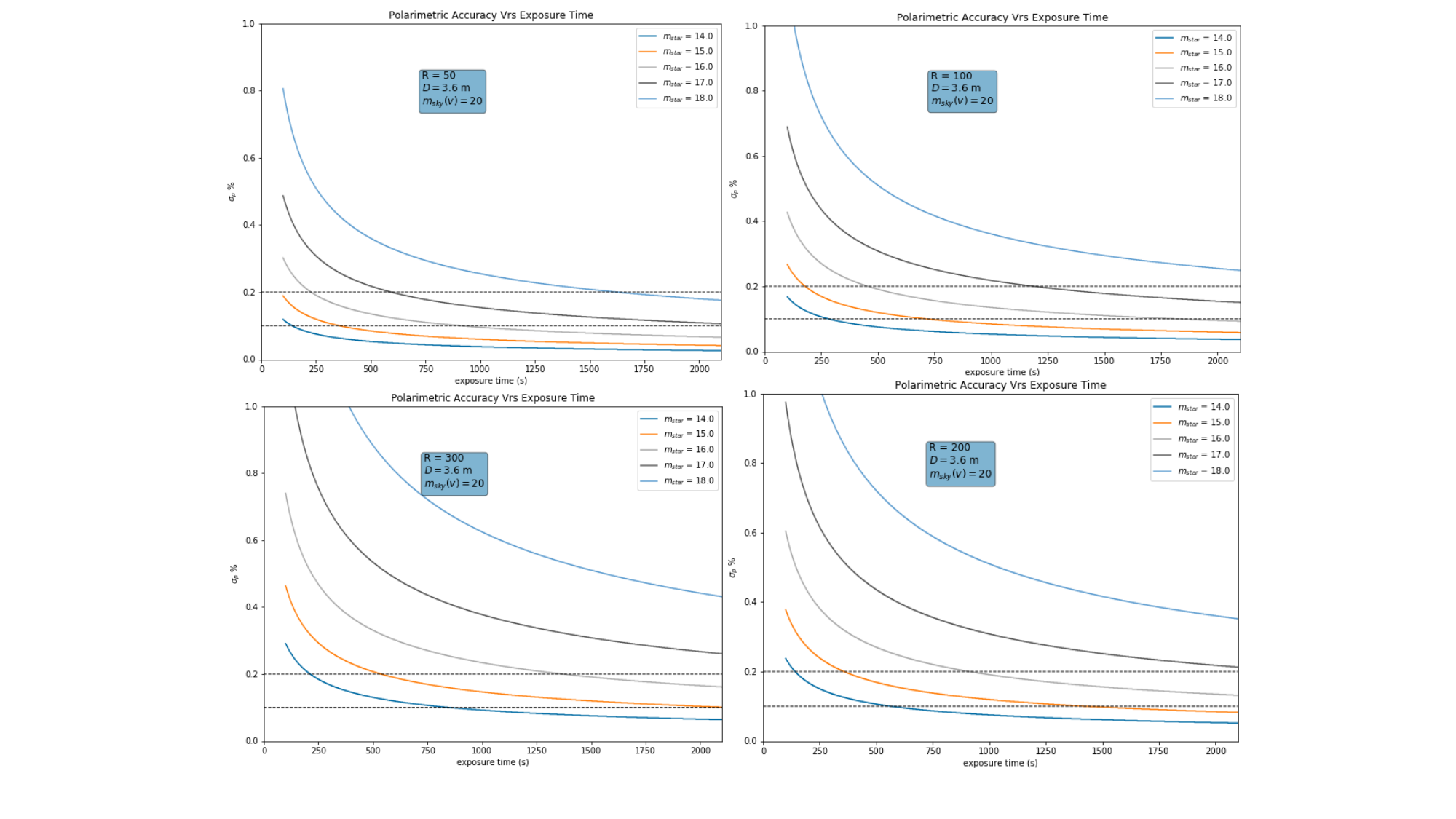}
\label{4m_exptime}
\centering
\end{figure}

\begin{figure}[t]
\centering
\caption{Polarimetric accuracy vs. exposure plots for different visual brightness of stars observed with 2m telescope.}
\includegraphics[trim = {5cm 0cm 6cm 0cm}, width=18cm]{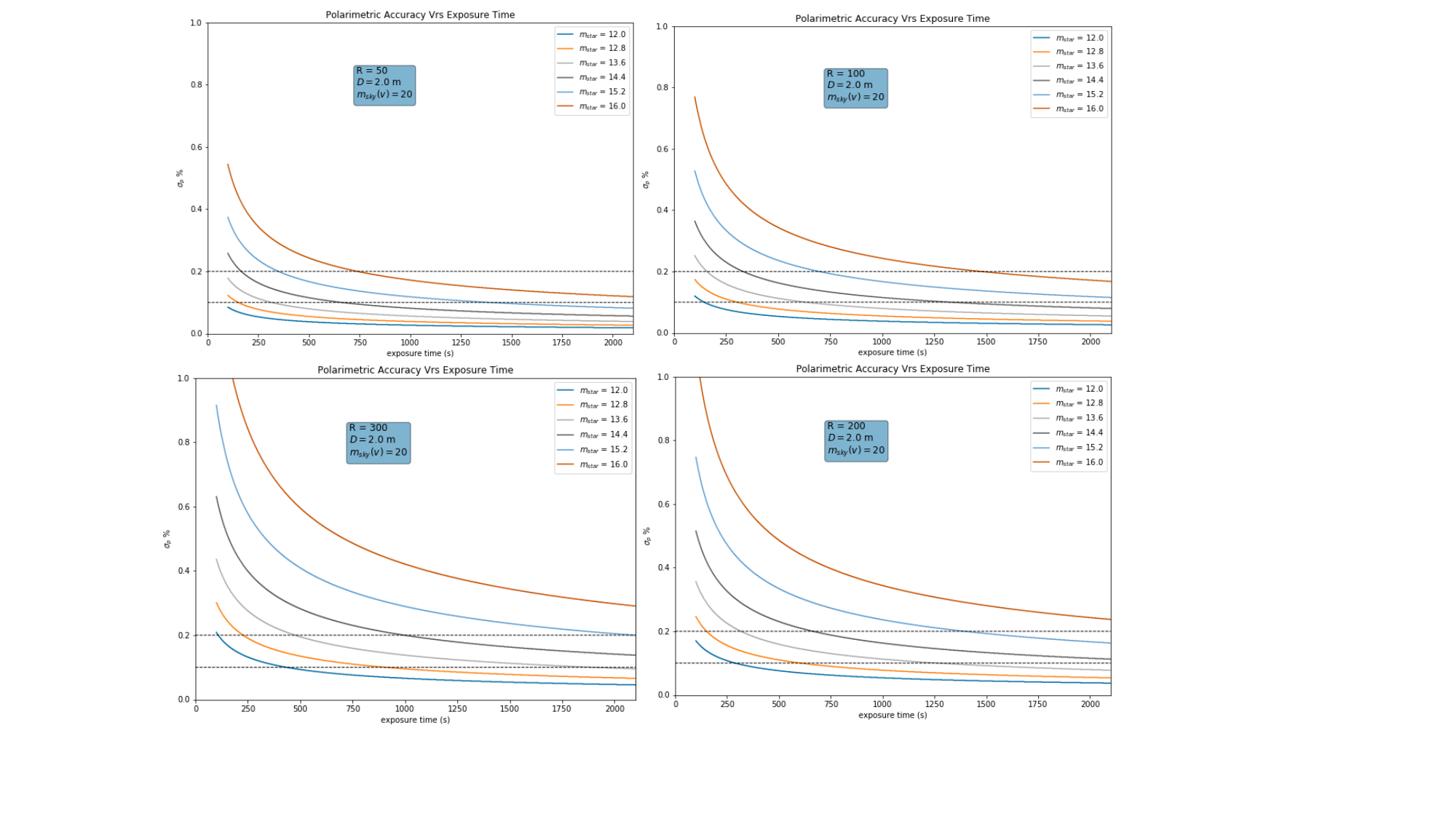}
\label{2m_exptime}
\end{figure}

As discussed above, in order to achieve an polarimetric accuracy of 0.1~\%, a photometric SNR of around 1400 is needed, which makes accurate polarimetry limited in sky depth. To reach within  0.1~\% accuracy for objects up to 18 mags in V band, a 4~m class telescope is needed. We have carried out calculations of the estimated polarimetric accuracy as a function of exposure time for various spectral resolutions. The atmospheric and site details used for the calculations are noted in Table~\ref{inst_table}. Figure~\ref{4m_exptime} plots the polarimetric accuracy for a 4~m telescope as a function of exposure time and stellar magnitude at different spectral resolutions centered at the wavelength of 5550~{\AA}. Figure~\ref{2m_exptime} plots the corresponding results for a 2~m telescope.

\par As mandated by the science cases described above, for the instrument, a FoV of 5-10 arcminutes is required. This proposed instrument will be a novel combination of wide-field, high accuracy spectropolarimetry and imaging polarimetry. Table~\ref{inst_spec_table} lists the desired technical specifications for the instrument. There are multiple design implementation possibilities for such an instrument, which we will briefly discuss below. 

\begin{table}[t]
    \centering
    \begin{tabular}{cc}
    \hline
    \textbf{Parameter}     & \textbf{Value}  \\
    \hline
      seeing Full-Width Half-Maximum  & 1 arcseconds \\
     sky brightness & 20 mags/arcseconds$^{2}$ in V band\\
     Camera f/\# & 4.5\\
      & Two-channel and two-camera \\
     Type of polarimetry & OR\\
     & four-channel and four-camera \\
     Instrument throughput & 50 ~\% \\
     \hline
    \end{tabular}
    \caption{details of the site and instrument in calculations of the exposure time versus accuracy.}
    \label{inst_table}
\end{table}

\pagebreak
\subsection{Possible instrument design concepts}
As the name suggests, a spectropolarimeter disperses light  along with splitting the orthogonal polarizations of the light from source. The architecture of the polarization analyzer unit will be driven by the technical requirements of the instrument. To develop a low-resolution spectropolarimeter, below are the possible ways to do design the analyzer system-  it will consist of dispersion and polarization analysis units for creating the spectra and separating polarizations, respectively:\\

{\bf Dispersion Element:}
\begin{enumerate}
    \item Grisms: We can use grisms to work as a dispersing element. This allows the flexibility for the grisms to be inserted in the filter wheel of a Faint Object Spectrograph and Camera (FOSC) instrument, enabling the use of multiple gratings and filters, allowing the capability of variable spectral resolution spectropolarimetery as well as imaging polarimetry.
    \item Glass Prisms eg. Amici Prism: These prisms allow for creating dispersion that can be controlled, and changed and yet, also be made achromatic (no-dispersion, to allow imaging polarimetry. The dispersion element can be placed after the Wollaston prisms (WPs) in the collimated beam path near pupil. Such elements are often used when a desired resolution (R) of the range of 50-200 is required. A major advantage of these prisms is that they can be easily integrated with existing FOSC like instruments by placing them in the filter wheel. 
    \item Polarization gratings: We can combine dispersion and polarization analyzer systems into one by using polarization gratings. As the name may indicate, these novel and upcoming gratings have the combined capabilities of separating the two circular polarizations while also acting as a dispersion element$^{27, 40}$; these elements can provide an accuracy of better than 0.1\% . As these separate the two circular polarization components, we will need a rotating quarter-wave plate upstream to measure linear polarization.

    \item Calcite Wollaston prisms: We can use readily available Wollaston prisms made of Calcite for a desired resolution of approximately 50. This approach is being attempted with the RSS instrument on SALT telescope$^{30}$.
\end{enumerate}

{\bf Polarization Analyzer:} Following options can be used to work as analyzer in a polarimeter. 
\begin{enumerate}
    \item Polarization gratings.
    
    \item Single/Double Wollaston Prisms. These are used ubiquitously in astronomy to obtain high accuracy polarimetric measurements$^{31, 32}$. 
    \item Wire-grid polarization beam-splitter: they have uniform behavior over large wavelengths. For example, the MOPTOP instrument uses this for imaging polarimetry$^{37}$. The choice of wire-grid may provide an accuracy of 0.2\% on the sky.
\end{enumerate}

\section{Calibration, Data reduction and polarization estimation methods}
Data reduction will be done using standard data reduction routines available to the community. Calibration of wide-field polarimetry instruments is a challenging problem owing to the combination of the following two effects: (a) for wide-field instruments, there is large change in the polarimetric response function of the instrument across the FoV, requiring wide-field calibration sources that can fill the entire FoV, and  (b) currently, prior to the work by the authors$^{22, 23}$, there have been no established wide-field calibration sources. A more detailed discussion on this subject can be found in the polarimetric modelling and calibration paper for WALOP instruments by Maharana et al (ref. 21). They have developed a high accuracy polarimetry calibration method for wide-field polarimeters, in particular for WALOP-South instrument.

\section{Current status and future}

The proposed instrument is currently under discussion phase. The team is meeting regularly and discussing the science cases, technical requirements, telescope size justification etc. We have completed the process of selecting suitable science cases for the instrument. The next steps include preparing a proposal for the development of this instrument and presenting it to the observatories in India. 
\\

{\bf References:}

\begin{enumerate}
    \item Alves, F. O., Frau, P., Girart, J. M. et al., On the radiation driven alignment of dust grains: Detection of the polarization hole in a starless core, $A\&A$, 2014, 569, L1.
    \item Andersson, B-G., Lazarian, A., \& Vaillancourt, J. E., Interstellar Dust Gain Alignment, \textit{Annual Review of Astronomy and Astrophysics}, 2015, 53, p. 501.
    \item Andersson, B-G., \& Potter, S. B., Observational Constraints on Interstellar Grain Alignment, $ApJ$, 2007, 665, p. 369.
    \item Andersson, B-G., Lopez-Rodriguez, E., Medan, I., et al., Grain Alignment in the Circumstellar Shell of IRC$+$10$^{\circ}$ 216, $ApJ$, 2022, 931, p. 80.
    \item Blinov, D., Maharana, S., Bouzelou, F., et al., The RoboPol sample of optical polarimetric standards, $A\&A$, 2023, 677, p. 144.
    \item Cashman, L. R., \& Clemens, D. P., THE MAGNETIC FIELD OF CLOUD 3 IN L204, $ApJ$, 2014, 793, p. 126.
    \item Clayton, G. C., Wolff, M. J., Allen, R. G., \& Lupie, O. L., Ultraviolet Interstellar Linear Polarization. II. The Wavelength Dependence, $ApJ$, 1995, 445, p. 947.
    \item Cohen, M., \& Schmidt, G. D., Spectropolarimetry of Herbig-Haro objects and the exciting star of HH 30, $AJ$, 1981, 86, p. 1228.
    \item Cotton, D. V., Marshall, Jonathan P. ; Frisch, Priscilla C. et al., The wavelength dependence of interstellar polarization in the Local Hot Bubble, $MNRAS$, 2019, 483, 3, p.3636-3646    
    \item Gonzalez-Gaitan, S., Mourao, A. M., Patat, F., et al., Tips and tricks in linear imaging polarimetry of extended sources with FORS2 at the VLT, $A\&A$, 2020, 634, p. 70.
    \item Hall, J. S., Observations of the Polarized Light From Stars, Science, 1949, 109, p. 166.
    \item Heiles, C., 9286 Stars: An Agglomeration of Stellar Polarization Catalogs, \textit{The Astronomical Journal}, 2000, 119, p. 923.
    \item Hensley, B. S., \& Draine, B. T., Detection of PAH Absorption and Determination of the Mid-infrared Diffuse Interstellar Extinction Curve from the Sight Line toward Cyg OB2-12, $ApJ$, 2020, 895, p. 38.
    \item Hiltner, W. A., Polarization of Light From Distant Stars by Interstellar Medium, $Science$, 1949, 109, p. 165.
    \item Hoang, T., Tram, L. N., Lee, H., \& Ahn, S.-H., Rotational disruption of dust grains by radiative torques in strong radiation fields, \textit{Nature Astronomy}, 2019, 3, p. 766.
    \item Jones, T. J., Bagley, M., Krejny, M., Andersson, B. G., \& Bastien, P., GRAIN ALIGNMENT IN STARLESS CORES, $AJ$, 2014,149, p. 31.
    \item Kim, S.-H., \& Martin, P. G., The Size Distribution of Interstellar Dust Particles as Determined from Polarization: Spheroids, $ApJ$, 1995, 444, 293.
    \item Lazarian, A., \& Hoang, T. Radiative torques: analytical model and basic properties, $MNRAS$, 2007, 378, p. 910.
    \item Maharana, S., Kypriotakis, J. A., Ramaprakash, A. N., et al. WALOP-South: A wide-field one-shot linear optical polarimeter for PASIPHAE survey in \textit{Ground-based and Airborne Instrumentation for Astronomy VIII}, ed. C. J. Evans, J. J. Bryant, \& K. Motohara, Vol. 11447, International Society for Optics and Photonics (SPIE), 2020, p.1135 – 1146.
    \item Maharana, S., Kypriotakis, J. A., Ramaprakash, A. N., et al., WALOP-South: a four-camera one-shot imaging polarimeter for PASIPHAE survey. Paper I—optical design, \textit{Journal of Astronomical Telescopes, Instruments}, and Systems, 2021, 7, p. 1.
    \item Maharana, S., Anche, R. M., Ramaprakash, A. N., et al., WALOP-South: a four-camera one-shot imaging polarimeter for PASIPHAE survey. Paper II – polarimetric modeling and calibration, \textit{Journal of Astronomical Telescopes, Instruments, and Systems}, 2022, 8, 038004.
    \item Maharana, S., Kiehlmann, S., Blinov, D., et al., Bright-Moon sky as a wide-field linear Polarimetric flat source for calibration, $A\&A$, 2023, 679, p. 68
    \item Mandarakas, N., Panopoulou, G. V., Pelgrims, V., et al., Zero-polarization candidate regions for calibration of wide-field optical polarimeters, Zero-polarization candidate regions for calibration of wide-field optical polarimeters, $A\&A$, 2024, DOI: https://doi.org/10.1051/0004-6361/202348099
    \item Mathis, J. S., Rumpl, W., \& Nordsieck, K. H., The size distribution of interstellar grains, $ApJ$, 1977, 217, p. 425.
    \item Medan, I., \& Andersson, B. G., Magnetic Field Strengths and Variations in Grain Alignment in the Local Bubble Wall, $ApJ$, 2019, 873, p. 87.
    \item Moran, E. C., Spectropolarimetry Surveys of Obscured AGNs in \textit{Astronomical Society of the Pacific Conference Series}, The Central Engine of Active Galactic Nuclei, ed. L. C. Ho \& J. W. Wang, 373, 2007, p. 425.
    \item Packham, C., Escuti, M., Ginn, J., et al., Polarization Gratings: A Novel Polarimetric Component for Astronomical Instruments, \textit{Publications of the Astronomical Society of the Pacific}, 2010, 122, p. 1471.
    \item Panopoulou, G. V., Hensley, B. S., Skalidis, R., Blinov, D., \& Tassis, K., Extreme starlight polarization in a region with highly polarized dust emission, $A\&A$, 2019, 624, L8.
    \item Patat, F., Taubenberger, S., Cox, N. L. J., et al., Properties of extragalactic dust inferred from linear polarimetry of Type Ia Supernovae, $A\&A$, 2015, 577, A53.
    \item Potter, S. B., Nordsieck, K., Romero-Colmenero, E., et al., Commissioning the polarimetric modes of the Robert Stobie spectrograph on the Southern African Large Telescope in \textit{Ground-based and Airborne Instrumentation for Astronomy VI}, ed. C. J. Evans, L. Simard, \& H. Takami, International Society for Optics and Photonics (SPIE), 9908, 2016, 810 – 817.
    \item Ramaprakash, A. N., Rajarshi, C. V., Das, H. K., et al., RoboPol: a four-channel optical imaging polarimeter, $MNRAS$, 2019, 485, p. 2355.
    \item Ramaprakash, A. N., Gupta, R., Sen, A. K., \& Tandon, S. N. An imaging polarimeter (IMPOL) for multi-wavelength observations, \textit{Astron. Astrophys. Suppl. Ser.}, 1998, 128, p. 369.
    \item Sandin, C., Three-component modelling of C-rich AGB star winds – IV. Revised interpretation with improved numerical descriptions, $MNRAS$, 2008, 385, p. 215.
    \item Schmidt, G. D., \& Miller, J. S., The emission/reflection nature of Herbig-Haro object 24, $ApJL$, 1979, 234, L191.
    \item Semenko, E., Romanyuk, I., Yakunin, I., Kudryavtsev, D., \& Moiseeva, A., Spectropolarimetry of magnetic Chemically Peculiar stars in the Orion OB1 association, $MNRAS$, 2022, 515, p. 998.
    \item Serkowski, K., \textit{Interstellar Dust and Related Topics}, ed. J. M. Greenberg \& H. C. van de Hulst, 1973, 52, p. 145.
    \item Shrestha, M., Steele, I. A., Piascik, A. S., et al., Characterization of a dual-beam, dual-camera optical imaging polarimeter, $MNRAS$, 2020, 494, p. 4676.
    \item Soam, A., Andersson, B. G., Straizys, V., et al., Interstellar Extinction, Polarization, and Grain Alignment in the Sh 2-185 (IC 59 and IC 63) Region , $AJ$, 2021, 161, p. 149.
    \item Tassis, K., Ramaprakash, A. N., Readhead, A. C. S., et al., PASIPHAE: A high-Galactic-latitude, high-accuracy optopolarimetric survey., 2018, \textit{https://arxiv.org/abs/1810.05652}.
    \item Tinyanont, S., Millar-Blanchaer, M. A., Nilsson, R., et al., WIRC+Pol: A Low-resolution Near-infrared Spectropolarimeter, \textit{Publications of the Astronomical Society of the Pacific}, 2018, 131, 025001.
    \item Vaillancourt, J. E., Andersson, B. G., Clemens, D. P., et al., Probing Interstellar Grain Growth through Polarimetry in the Taurus Cloud Complex, $ApJ$, 2020, 905, p. 157.
    \item Whittet, D. C. B., Gerakines, P. A., Hough, J. H., \& Shenoy, S. S., Interstellar Extinction and Polarization in the Taurus Dark Clouds: The Optical Properties of Dust near the Diffuse/Dense Cloud Interface, $ApJ$, 2001, 547, p. 872. 
    \item Zejmo, M., Slowikowska, A., Krzeszowski, K., Reig, P., \& Blinov, D., Optical linear polarization of 74 white dwarfs with the RoboPol polarimeter, $MNRAS$, 2017, 464, p. 1294.
\end{enumerate}




\end{document}